\documentclass[letterpaper,12pt,onecolumn]{article}
\usepackage[T1]{fontenc}
\usepackage[utf8]{inputenc}
\usepackage{authblk}
\usepackage{graphicx}
\usepackage{lscape}
\usepackage{soul}
\usepackage{lineno}
\usepackage{amsmath,amssymb,latexsym}

\begin{document}

\title{\textbf{Titan  brighter at twilight than in daylight}}

\author[1]{A. Garc\'ia Mu\~noz \footnote{Corresponding author: garciamunoz@astro.physik.tu-berlin.de; tonhingm@gmail.com}}
\affil[1]{Zentrum f\"ur Astronomie und Astrophysik, Technische Universit\"at Berlin, D-10623 Berlin, Germany}

\author[2]{P. Lavvas}
\affil[2]{Groupe de Spectrom\'etrie Mol\'eculaire et Atmosph\'erique, 
UMR 7331, CNRS, Université de Reims Champagne-Ardenne, Reims 51687, France}

\author[3]{R.A. West}
\affil[3]{Jet Propulsion Laboratory, California Institute of Technology, 4800 Oak Grove Drive, Pasadena, California 91109, USA}


\maketitle

\newpage

\begin{centering}

Pre-print of a manuscript published in \\
NATURE ASTRONOMY 1, 0114 (2017) \\ 
DOI: 10.1038/s41550-017-0114  \\
www.nature.com/nastronomy \\
\end{centering}

\newpage

\textbf{
Investigating the overall brightness of planets (and moons) 
provides insight into their envelopes and energy budgets  
\cite{russell1916,arkingpotter1968,goodeetal2001,mallamaetal2002}. 
Titan phase curves (a representation of overall brightness vs. Sun-object-observer phase angle) 
have been published over a limited range of phase angles and spectral passbands 
\cite{tomaskosmith1982,westetal1983}.
Such information has been key to the study of the stratification, 
microphysics and aggregate nature of Titan's atmospheric haze 
\cite{ragespollack1983,westsmith1991}, 
and has complemented the spatially-resolved observations first showing  
that the haze scatters efficiently in the forward direction 
\cite{ragespollack1983,ragesetal1983}. 
Here we present Cassini Imaging Science Subsystem whole-disk brightness measurements of Titan 
from ultraviolet to near-infrared wavelengths. 
The observations reveal that Titan's twilight 
(loosely defined as the view when the phase angle $\gtrsim$150$^{\circ}$) 
outshines its daylight at various wavelengths. 
From the match between measurements and models, 
we show that at even larger phase angles the back-illuminated moon will appear much
brighter than when fully illuminated. This behavior is unique to Titan in our solar system, 
and is caused by its extended atmosphere and the efficient forward scattering of sunlight 
by its atmospheric haze.  
We infer 
a solar energy deposition rate (for a solar constant of 14.9 Wm$^{-2}$) of 
(2.84$\pm$0.11)$\times$10$^{14}$ W, 
consistent to within 1-2 standard deviations with Titan's time-varying thermal emission spanning 2007-2013  
\cite{lietal2011,li2015}. 
We propose that a forward scattering signature may also occur at large phase angles 
in the brightness of exoplanets with extended hazy atmospheres, 
and that this signature has valuable diagnostic potential for atmospheric characterization.
}

\newpage

We produced Titan phase curves (Fig. 1) from 
calibrated, whole-disk images taken with the Narrow Angle Camera of the 
Cassini Imaging Science Subsystem (ISS) \cite{porcoetal2004,westetal2010}.  
The timespan of the images (2004--2015), 
phase angle coverage ($\alpha$$\le$166$^{\circ}$; 
see Supplementary Fig. 1 for sketch of the viewing geometry), 
sampling ($\sim$400 datapoints/curve on average), 
and variety of filters 
(15, effective wavelengths $\lambda_{\rm{eff}}$=300--940 nm) 
in this work significantly expand on previous treatments
\cite{tomaskosmith1982,westetal1983}. 
The phase curves are presented in the size-normalized way $A_{\rm{g}}\Phi$($\alpha$), 
adopting Titan's solid radius of 2575 km
for the normalization.
$A_{\rm{g}}$ is the geometric albedo  and 
$\Phi$($\alpha$=0)$\equiv$1 by definition. 

From full illumination ($\alpha$=0) to $\alpha$$\sim$90$^{\circ}$, 
the curves describe Titan's progressive dimming as less
of its dayside appears visible to the observer. Our Cassini/ISS data
confirm 
the Pioneer 11 measurements in blue and red passbands
for $\alpha$$\leq$96$^{\circ}$  \cite{tomaskosmith1982} 
(Fig. 1)
 and ground-based spectroscopic data at $\alpha$$\sim$5.7$^{\circ}$ 
\cite{karkoschka1998} (Fig. 2).  
The incomplete sampling near full illumination does not permit the confirmation of 
proposed strong backscattering at very small phase angles and short wavelengths \cite{tomaskoetal2008,dooseetal2016}, 
a task that
would require nearly continuous sampling for $\alpha$$\le$5$^{\circ}$ in the passbands with 
$\lambda_{\rm{eff}}$=441 and 455 nm (Fig. 1). We note though that the data for
$\lambda_{\rm{eff}}$=343 nm do not show a discernible enhancement in the antisolar direction, 
which suggests that
backscattering in the ultraviolet is less strong than recently proposed.
The curve for the narrowband filter with $\lambda_{\rm{eff}}$=938 nm is particularly
well sampled with nearly 2000 datapoints. In this passband, Titan's surface contributes to the
emergent radiation (Fig. 3). 
The dispersion in the $A_{\rm{g}}\Phi$($\alpha$) values for this passband
at small phase angles (probing deeper in the atmosphere) 
is tentatively attributed to rotational effects as different surface patches 
enter and exit the field of view  
\cite{lemmonetal1995, negraoetal2006}.

For larger phase angles up to $\sim$140$^{\circ}$, 
the diminishing size of Titan's visible dayside 
is compensated by efficient forward scattering of its upper-atmosphere haze, 
and the curves exhibit plateaus or mild increases in brightness. 
As Titan nears back-illumination ($\alpha$$\geq$150$^{\circ}$), 
the twilight brightness rises steeply 
and at $\alpha$$\sim$160--166$^{\circ}$ it is 
comparable to or higher than at full illumination. 
This is particularly the case for wavelengths at which absorption by haze ($\lesssim$600 nm) 
or atmospheric methane (strong bands occur
at e.g. 619, 727, and 889 nm; Fig. 2) make Titan's dayside darker.
The brightness surge is unrelated to the central flash caused by atmospheric refraction 
during stellar occultations \cite{hubbardetal1993}.

We have investigated the empirical phase curves with a Monte Carlo 
radiative transfer model that solves the multiple-scattering problem of 
reflected sunlight in stratified, spherical-shell atmospheres 
\cite{garciamunozmills2015} (`Methods').
Our implementation of aerosol optical properties 
(extinction coefficients, single scattering albedos, scattering phase functions) 
is based on the latest interpretation of the in-situ measurements made 
by the Huygens Descent Imager/Spectral Radiometer (DISR) \cite{dooseetal2016}.
The DISR observations were made from 150 km altitude down to the surface, but
they were sensitive to aerosols above 150 km, as confirmed by this study.

The model phase curves based on the DISR aerosol implementation
(solid red curves, Fig. 1) reproduce well the Cassini/ISS data
for wavelengths $\ge$440 nm unaffected by methane absorption. 
The DISR implementation is poorly constrained shortwards of 490 nm though
\cite{tomaskoetal2008,dooseetal2016}. 
To reproduce the observations in the passbands with $\lambda_{\rm{eff}}$=306 and 343 nm,
we slightly modified the aerosol absorption above 150 km, which 
led to better fits (dashed red curves) (`Methods'). 
In the passbands affected by methane, an adjustable amount of methane 
absorption at all altitudes provided the required attenuation to reproduce the 
observations (dashed red curves, $\lambda_{\rm{eff}}$$\ge$619 nm).
The reported best fits represented by the dashed red curves were the result of minimizing the relative
measurement-model error.

The grey areas of Fig. 1
quantify the amount of energy that Titan scatters in all possible three-dimensional 
directions with a phase angle $\alpha$. This is mathematically expressed by
$\Phi$($\alpha$)$\sin$($\alpha$) \cite{russell1916}. 
The integrated area (properly normalized) is the phase integral $q$, 
and depends on wavelength (`Methods'). 
Thus, our inferred phase integrals are passband-averaged values. 
According to our best-fitting models, phase angles $\alpha$$\ge$166$^{\circ}$ 
contribute to $q$ by 13\% ($\lambda_{\rm{eff}}$=343 nm), 5\% (649 nm) and 7\% (928 nm). 
These non-negligible contributions originate from layers above 150 km (Fig. 3)
and substantiate the role of the upper-atmosphere haze in Titan's energy balance 
by scattering part of the incident sunlight. 
The phase integrals calculated here ($q$=1.9--2.9) are notably larger than earlier
estimates ($q$=1.3--1.7) based on incomplete phase angle coverage 
\cite{tomaskosmith1982,younkin1974}. 

At small phase angles, Titan's brightness is dictated by solar photons 
that scatter tens of times  before exiting the atmosphere. 
Our model shows that 
the photons emerging from the $\alpha$=0 configuration scatter preferentially
at altitudes of 150--300 km (455-nm photons) and 50--250 km (938-nm photons) 
(Fig. 3).
In contrast, the brightness for large phase angles is caused by photons
undergoing only a few collisions. 
For 455- and 938-nm photons and $\alpha$=166$^{\circ}$, 
the preferential range of scattering altitudes is 250--350 km and 200--350 km, respectively.
As a rule, the larger phase angles are more sensitive to higher altitudes. 

Atmospheric stratification is key in the interpretation of the twilight brightness, 
and the product $p_{\rm{a}}$($\theta$$\rightarrow$0)$H_{\rm{a}}/R$
of the particles scattering phase function ($\theta$ is the scattering angle between
the incident and exiting photon directions), 
$p_{\rm{a}}$($\theta$), times the ratio of the aerosol scale height to Titan's radius,
$H_{\rm{a}}/R$, becomes a key factor. 
Titan's haze particles are fractal aggregates, each comprising 
thousands of small ($\le$0.05 $\mu$m) spherical monomers  
\cite{westsmith1991,lavvasetal2010}.
To the effects of radiative transfer modeling, 
the haze particles behave with their own aggregate-averaged properties, which 
may significantly differ from those of the monomers.
The large effective size of the aggregates
(equal-projected-area radii of 2--3 $\mu$m \cite{tomaskoetal2008,lavvasetal2010, rannouetal2010})
causes the prominent forward lobe in the particles scattering phase function that has 
been known since the times of the Voyager spacecraft \cite{ragespollack1983,ragesetal1983}
and that translates into a large $p_{\rm{a}}$($\theta$$\rightarrow$0).
The DISR measurements, some of them obtained while looking less than 10$^{\circ}$ away 
from the Sun, have shown that the forward lobe had been severely underestimated 
for decades \cite{tomaskoetal2008,dooseetal2016}.
Titan's extended atmosphere results in 
$H_{\rm{a}}/R$$\sim$45/3000$\sim$1.5$\times$10$^{-2}$, 
much larger than e.g. $\sim$4/6150$\sim$6.5$\times$10$^{-4}$ and
$\sim$27/70000$\sim$3.8$\times$10$^{-4}$ for Venus and Jupiter, respectively.
This difference has an impact on their corresponding forward scattering components
\cite{garciamunozmills2015, mallamaetal2006}.

The large value of $p_{\rm{a}}$($\theta$$\rightarrow$0)$H_{\rm{a}}/R$ in Titan's upper atmosphere 
is ultimately responsible for the whole-disk brightness surge at large phase angles.
Based on our model's capacity to reproduce the 
measured phase curves for $\alpha$$\le$166$^{\circ}$, 
we predict that Titan's brightness for $\alpha$$\rightarrow$180$^{\circ}$ 
exceeds the full-illumination brightness by an order of magnitude or more, 
depending on the observation wavelength. 
The predicted $A_{\rm{g}}$$\Phi$($\alpha$$\rightarrow$180$^{\circ}$) 
are quoted in Fig. 1. 
The diminishment in particle sizes (and in the strength of their forward scattering lobe) 
above 400 km  \cite{lavvasetal2010}
has no impact on the brightness surge because the atmosphere at these altitudes is 
optically thin at the wavelengths investigated here.

The Titan aerosols participate in vertical and horizontal structures, 
and in transient behaviors on diverse timescales \cite{tomaskowest2009,westetal2014}. 
This complexity results from the strong ties of aerosol formation 
with super-rotating winds, seasonal heating, and both
neutral and ion photochemistry. 
The match 
between the Cassini/ISS phase curves and models over all measured phase
angles supports 
the DISR implementation \cite{dooseetal2016} as a functional 
representation of the globally-averaged optical properties of Titan's aerosols.
This conclusion was unanticipated because Titan's brightness at the larger
phase angles is dictated by the atmosphere above 150 km
(Fig. 3), and also because in principle the DISR conclusions apply principally to
the Huygens descent conditions.

We have used the inferred whole-disk scattering properties to 
calculate the rate of solar energy deposited into the Titan atmosphere as the
difference between the incident and scattered rates, 
$P_{\rm{dep}}$=$P_{\rm{inc}}$$-$$P_{\rm{sca}}$ (`Methods').
For the incident contribution, scaled to a solar constant $S_{\odot}$=14.9 Wm$^{-2}$ 
specific to Saturn's semi-major axis of 9.58 AU, 
we obtain $P_{\rm{inc}}$=(3.87$\pm$0.07)$\times$10$^{14}$ W.
For the rate of energy scattered by Titan, 
we obtain $P_{\rm{sca}}$=(1.03$\pm$0.08)$\times$10$^{14}$ W, its uncertainty being
comparable to that for $P_{\rm{inc}}$. 
Figure 4 and Supplementary Table 1
summarize the partial contributions to both
$P_{\rm{inc}}$ and $P_{\rm{sca}}$. 
Finally, we infer $P_{\rm{dep}}$=(2.84$\pm$0.11)$\times$10$^{14}$ W and,  
from the definition of Bond albedo, 
$A_{\rm{B}}$=$P_{\rm{sca}}$/$P_{\rm{in}}$=0.27$\pm$0.02. This is strikingly
similar to previous estimates of the Bond albedo \cite{tomaskosmith1982,younkin1974,neffetal1985} 
even though some of the intermediate quantities used in these works (including the phase integrals) 
were poorly constrained. 

Titan has been estimated to emit thermally at rates of
$P_{\rm{emi}}$=(2.86$\pm$0.01)$\times$10$^{14}$ W in 2007
($S_{\odot}$$\sim$16 Wm$^{-2}$) and
(2.79$\pm$0.01)$\times$10$^{14}$ W in 2013
($S_{\odot}$$\sim$14 Wm$^{-2}$) \cite{lietal2011,li2015}, 
suggesting that the emitted energy dropped less than the solar irradiation during that period.
Scaling $P_{\rm{dep}}$ from our time-averaged 
treatment of $P_{\rm{inc}}$ and $P_{\rm{sca}}$ by the relevant solar constants, we obtain 
$P_{\rm{dep}}$=(3.05$\pm$0.11)$\times$10$^{14}$ W in 2007
and (2.67$\pm$0.11)$\times$10$^{14}$ W in 2013.
Thus, $P_{\rm{dep}}$ and $P_{\rm{emi}}$ are consistent to within 1--2 standard deviations 
during 2007-2013. 
Our study cannot rule out an energy imbalance, but it sets strict limits based on
contemporaneous measurements of scattered sunlight and Titan's thermal emission. 
We note the order-of-magnitude difference in the uncertainties quoted for $P_{\rm{dep}}$ 
and $P_{\rm{emi}}$, and the 
difficulty of further constraining a putative imbalance if either the optical radius 
(a measure of Titan's sunlight-intercepting cross section, `Methods') 
or the wavelength-dependent reflectance change over time.

Titan’s brightness surge has direct implications on the characterization of exoplanets 
with extended and hazy atmospheres, two oft-cited properties amongst known 
exoplanets \cite{demoryseager2011,garciamunozisaak2015,lammeretal2016,singetal2016}. 
For illustration, 
we estimate (assuming a hydrogen-helium bulk composition 
and equilibrium temperature of 930 K) 
the background scale height of the low-gravity hot sub-Neptune CoRoT-24b to be 
$H$/$R$$\sim$3.5$\times$10$^{-2}$, and therefore larger than Titan’s $H_{\rm{a}}$/$R$.
Whether the haze on these exoplanets (provided it exists) produces significant forward scattering
is difficult to anticipate, as
current haze formation models 
\cite{hellingfomins2013,marleyetal2013} 
have limited capacities to predict particle sizes and the
corresponding $p_{\rm{a}}$($\theta$$\rightarrow$0).

Therefore, it remains plausible that some exoplanets
will exhibit brightness surges at large phase angles such as that experienced by Titan, 
which may in addition affect the measured transit radius \cite{dekokstam2012}.
Indeed, for a typical hot Jupiter on an edge-on orbit 
around a Sun-like star, phase angles of $\sim$175$^{\circ}$ are probed immediately 
before and after transit. 
In that viewing configuration, an extended atmosphere bearing Titan-like
haze will appear a few times brighter than indicated by its geometric albedo. 
The eventual detection of this phenomenon will help constrain the scale height and particle size of 
their atmospheric haze. 
Future theoretical work must therefore assess whether forward-scattering haze forms
in exoplanet atmospheres. Also, characterization efforts with existent data from the
CoRoT and Kepler missions, and with data from upcoming spacecraft such as CHEOPS, JWST, PLATO and TESS,  
should study the larger phase angles, as they offer diagnostic possibilities 
complementary to those usually explored with primary and secondary eclipses. 
 
\newpage

\cleardoublepage
\newpage

\newpage

\cleardoublepage
\newpage

\bibliographystyle{unsrt}

\newpage

\textbf{
Author to whom correspondence and requests for materials should be addressed:}

A. Garc\'ia Mu\~noz (garciamunoz@astro.physik.tu-berlin.de; tonhingm@gmail.com).
\\
\\
\\

\textbf{Acknowledgments}

A.G.M. gratefully acknowledges correspondence with L.A. Sromovsky 
and P.M. Fry on Voyager 2 observations. 
P.L. acknowledges financial support from the  Programme National de Plan\'etologie (PNP) of the INSU/CNRS.
\\
\\
\\

\textbf{Author contributions}

A.G.M. devised the research, performed the data reduction and model simulations, 
and wrote the manuscript.
P.L. provided various haze properties. R.A.W. provided insight into the treatment of images.
P.L. and R.A.W. provided valuable expertise in Titan's atmosphere. 
All authors discussed the content of the manuscript.
\\
\\
\\
\\

\newpage

   Figure 1.
   \textbf{Titan phase curves inferred in this work}. 
   Cassini/ISS measurements (black symbols) and model calculations (red curves; see text
   for meaning of solid and dashed lines). Each graph contains information on 
   filter combination and effective wavelength ($\lambda_{\rm{eff}}$), 
   mean relative difference between measurements and model ($\sigma$), 
   adopted single scattering albedo of the gas ($\omega_{0,\rm{g}}$), limiting values of
   $A_{\rm{g}}$$\Phi$($\alpha$) for $\alpha$$\rightarrow$0 and 180$^{\circ}$, and
   passband-averaged phase integral ($q$). The grey area shows (arbitrary normalization) 
   the product $\Phi$($\alpha$)$\sin{(\alpha)}$ that enters into 
   the evaluation of the phase integral $q$ ($\equiv$2$\int_{0}^{\pi}$$\Phi$($\alpha$)$\sin$($\alpha$)d$\alpha$). 
   Pioneer 11 phase curves in blue (452 nm) and red (648 nm) passbands 
   \cite{tomaskosmith1982} are shown in cyan color
   together with the Cassini/ISS curves for $\lambda_{\rm{eff}}$=455 and 649 nm. 
   The Pioneer 11 curves were 
   re-normalized from $R_{\rm{452nm}}$=2850 km and $R_{\rm{648nm}}$=2800 km to $R$=2575 km.
\\

   Figure 2.
   \textbf{Full-disk albedo spectrum at phase angle $\alpha$=5.7$^{\circ}$}. 
   Measurements from Earth (solid curve) \cite{karkoschka1998}, 
   and Cassini/ISS-based values for $\alpha$=5.7$^{\circ}$ interpolated from the 
   best-fits of Fig. 1 (symbols). 
   Each color symbol is matched by a color curve at the bottom. 
   Discontinuous curves represent the transmissions (arbitrary scale) for the 
   broad- and mediumband ISS filters; solid curves (arbitrary scale) refer to narrowband filters. 
\\

   Figure 3.
   \textbf{Radiative transfer modeling.}
   \textbf{a, b, }
   Cassini/ISS images of Titan at $\alpha$$\approx$0 
   and $\alpha$$\approx$166$^{\circ}$  obtained with the 
   CL1$\_$CB3 filter combination ($\lambda_{\rm{eff}}$=938 nm). 
   The contrast in the full-disk  image arises near the surface.
   The overall brightness of both configurations is nearly the same (Fig. 1), 
   despite the disparate
   projected area of Titan's illuminated disk in each case 
   (full disk for $\alpha$$\approx$0; 
   the equivalent to a ring a few scale heights wide for $\alpha$$\approx$166$^{\circ}$). 
   \textbf{c, d, }
   Contribution by altitude to overall brightness for model calculations specific to
   four filter combinations. This is the altitude where simulated 
   solar photons undergo their first scattering collision. 
   The contribution functions span 200--300 km (a few scale heights) and peak at
   higher altitudes for the larger phase angles. 
   \textbf{e, f, } 
   Contribution to overall brightness by number of scattering collisions undergone by the
   simulated photons. Same color code as in 
   \textbf{c, d}. The summation of the histogram over number of collisions 
   results in the measured $A_{\rm{g}}\Phi$($\alpha$) values. 
   Numerous collisions contribute to the brightness for small phase angles, 
   but only a few effectively contribute for large phase angles.
\\

 Figure 4. \textbf{
 Rates for incident solar energy and energy scattered by Titan}.  
 $F_{\odot}$($\lambda$) is the solar 
 irradiance and $R_{\rm{opt}}$($\lambda$) is Titan's optical radius. 
 $\pi$$R_{\rm{opt}}^2$($\lambda$)
 is the opaque area that would block the same amount of sunlight photons as Titan. 
 \textit{Normalized} quantities are 
   arbitrarily scaled. The solid area bracketing the $q$($\lambda$) curve conveys the
   adopted uncertainties in the phase integrals.    
   Dashed curves 
   represent the cumulative integrals leading to $P_{\rm{inc}}$ and
   $P_{\rm{sca}}$, respectively. See `Methods' for details.

   \begin{figure}[h]
   \centering
   \includegraphics[width=15.cm]{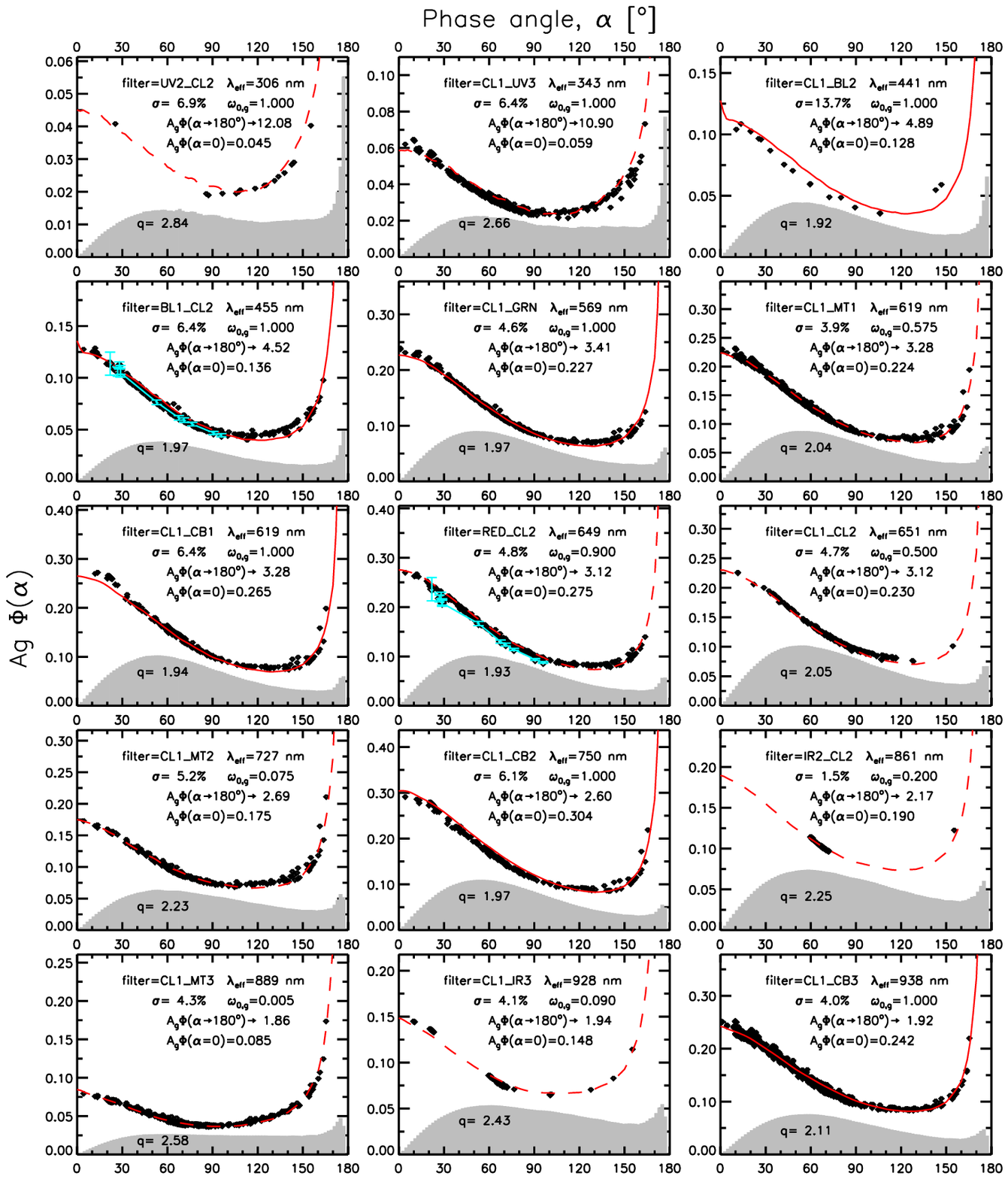}
   \caption{\label{mypanel_fig}
}
   \end{figure}

   \begin{figure}[h]
   \centering
   \includegraphics[width=9.cm]{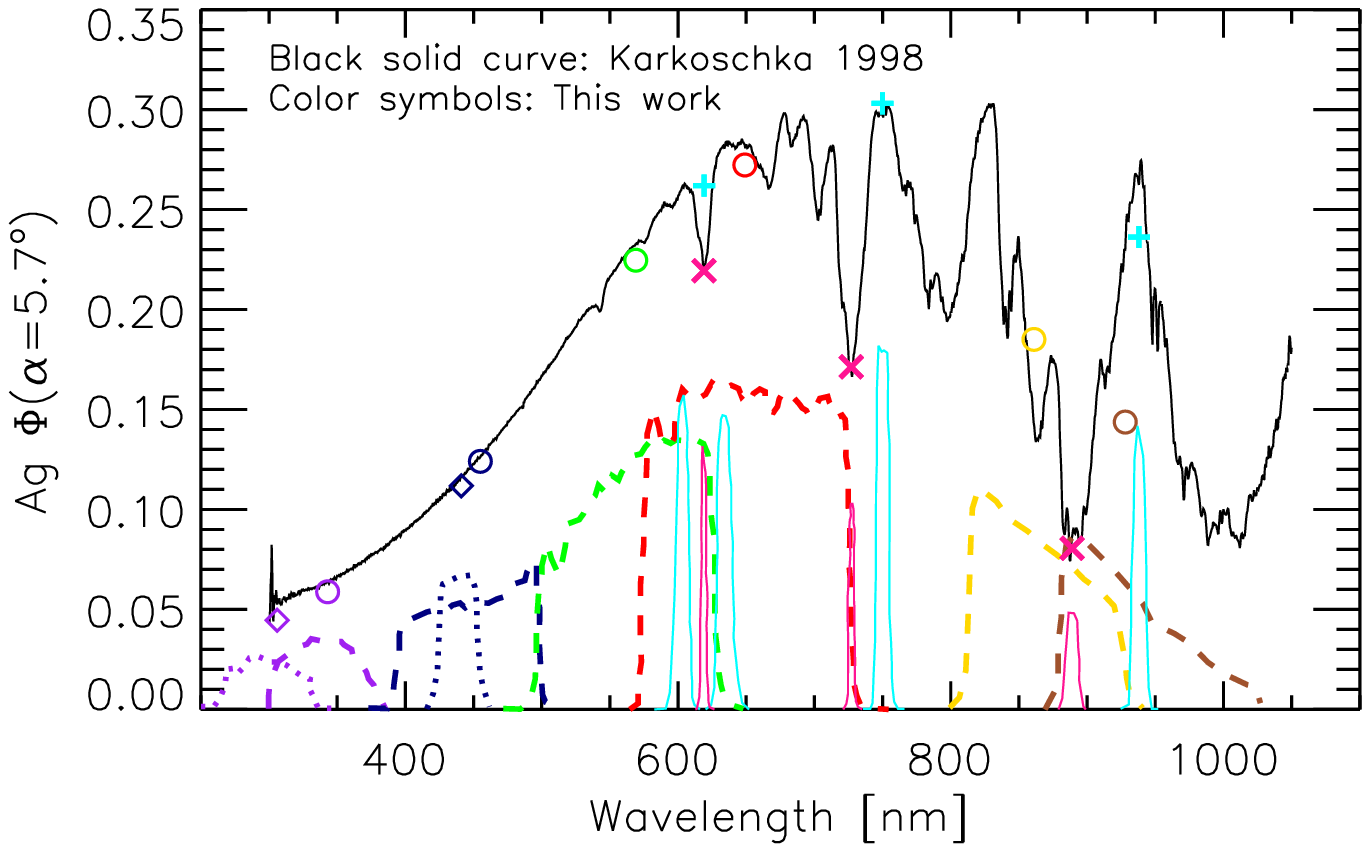}
   \caption{\label{myspectrum_fig} 
   }
   \end{figure}

   \begin{figure}[h]
   \centering
   \includegraphics[width=4.cm, height=4.cm]{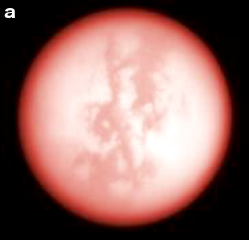}
   \includegraphics[width=4.cm, height=4.cm]{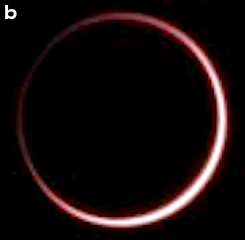}\\   
   \includegraphics[width=4.cm]{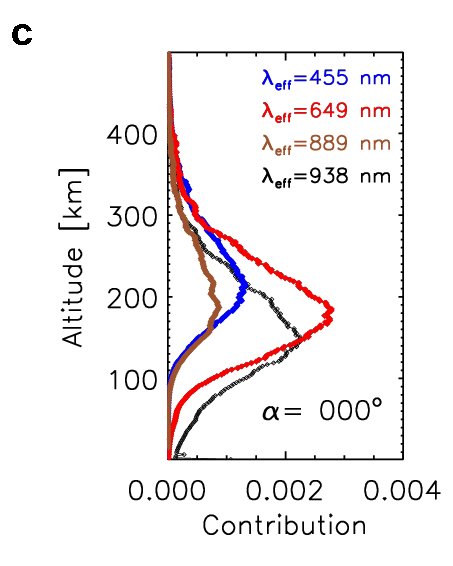}
   \includegraphics[width=4.cm]{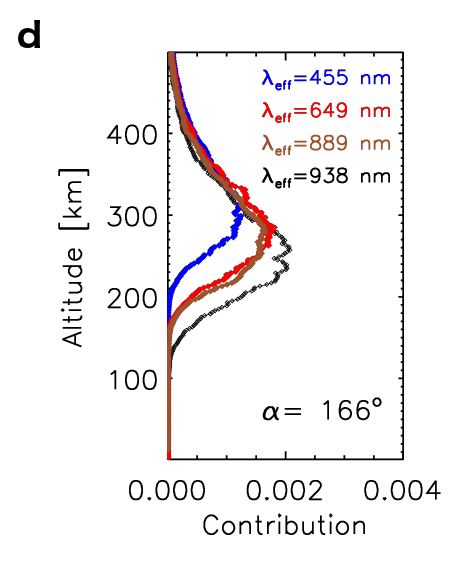}\\      
   \includegraphics[width=4.cm]{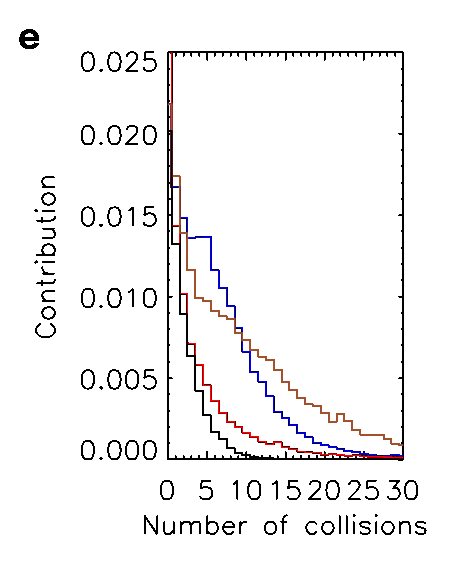}
   \includegraphics[width=4.cm]{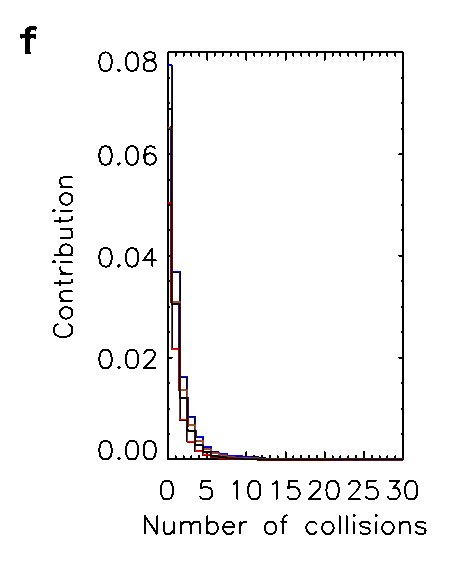}\\   
   \caption{\label{mydisk_fig} 
   }
   \end{figure}

   \begin{figure}[h]
   \centering
   \includegraphics[width=8.cm]{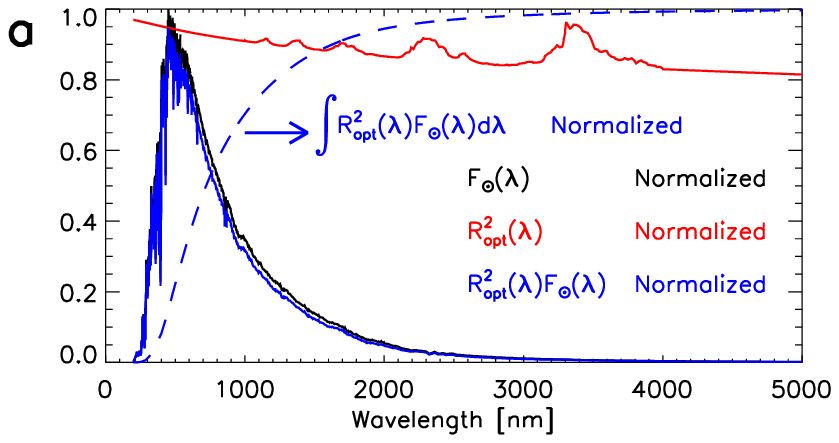}\\
   \includegraphics[width=8.cm]{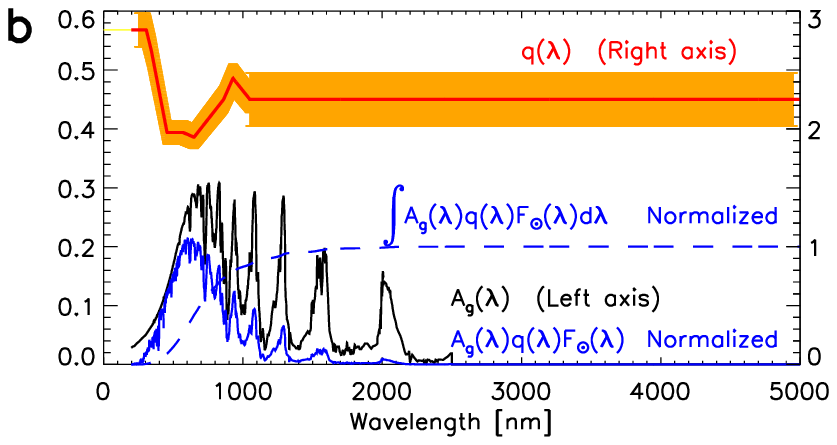}   
   \caption{\label{mybalance_fig}  
   }
   \end{figure}

\appendix
\cleardoublepage
\newpage

\textbf{\large{Methods}}\\

\textbf{Treatment of images}\\

Our search for whole-disk images of Titan utilized the 
Planetary Image Atlas (http://pds-imaging.jpl.nasa.gov/search/) and the
OPUS Data Search Tool at the Ring-Moon Systems Node of NASA's Planetary Data System 
(http://pds-rings.seti.org).
We focused on the Narrow Angle Camera, which is less susceptible to stray light
at large phase angles than the Wide Angle Camera [13, 35].
We identified 15 filter combinations for which there was a sufficient number of images. 
These filter combinations are characterized by effective wavelengths between
306 and 938 nm. Our search excluded polarization filters, although it will be very
interesting to combine in the future the brightness phase curves reported here with
polarization phase curves.
\\

We downloaded the photometrically-calibrated ($I/F$) images from 
the Ring-Moon System Node of NASA
(http://pds-rings.seti.org). Ancillary information such as phase angle was read from
the image headers. 
All the images were taken from farther than 1.2 million km away from Titan. 
The size of Titan in these images ranges from tens to a few hundreds of pixels.
We visually inspected each image, deselecting those that
were either truncated, damaged or overexposed. We identified a total of 5766 useable images,
taken between 2004 and 2015. The number of images per filter combination is:
UV2$\_$CL2 (16), 
CL1$\_$UV3 (422), 
CL1$\_$BL2 (19), 
BL1$\_$CL2 (353), 
CL1$\_$GRN (327), 
CL1$\_$MT1 (721), 
CL1$\_$CB1 (237), 
RED$\_$CL2 (314), 
CL1$\_$CL2 (238), 
CL1$\_$MT2 (286), 
CL1$\_$CB2 (273), 
IR2$\_$CL2 (73), 
CL1$\_$MT3 (464), 
CL1$\_$IR3 (121), 
CL1$\_$CB3 (1902).
\\

We conducted aperture photometry of Titan's whole disk in a standard way 
(e.g. [36]).
We drew on each image a circle of radius $r_{\rm{aper}}$=3500 km + 10 pixels 
from Titan's center and added the flux inside. 
We experimented with other radii to confirm that the 
point spread functions of the instrument-plus-filter optical systems 
were well contained within the aperture, and confirmed that larger aperture
radii did not affect the photometry by more than 1\%.
Some images were affected by a residual `dark current' 
that was estimated from a 10-pixel-wide annulus concentric with the aperture circle. 
Removing this residual component meant a reduction in the aperture photometry of 
2\% or less.
\\

If $I_j/F$ is the radiance from an image pixel $j$ [35],
 normalized to the incident solar flux $\pi$$F$ (dependent on the Sun-Titan distance), 
then the irradiance from the whole Titan disk at the spacecraft-Titan 
distance $\ell$ is:
\begin{equation}
\mathcal{F}=\sum_j (I_j/F) S_{\rm{pixel}} / \ell^2,
\label{eq1_eq}
\end{equation}
where the summation runs over all pixels $j$ within the circle of radius $r_{\rm{aper}}$,
and $S_{\rm{pixel}}$ is the projected area of a pixel at the distance $\ell$. 
$\mathcal{F}$ depends on the spacecraft-Titan distance, 
and is therefore of limited usefulness. 
Introducing:
\begin{equation} 
\mathcal{F}=\pi R^2/ \ell^2 A_{\rm{g}} \Phi(\alpha)
\label{eq2_eq}
\end{equation}
provides a size-normalized definition of Titan's overall reflectance, $A_{\rm{g}}$$\Phi(\alpha)$,
which can be evaluated through the combination of Eqs. (\ref{eq1_eq})--(\ref{eq2_eq}).
$A_{\rm{g}}$ is the geometric albedo and  $\Phi(\alpha)$ is the phase law.
There is no unique criterion to select the normalization radius $R$ in Eq. (\ref{eq2_eq}), 
and we adopted Titan's solid radius of 2575 km. 
Stating the selected normalization radius is critical to compare between works. \\

The amount of Cassini/ISS brightness measurements reported here and variety of conditions 
in which the observations were made will foreseeably trigger new investigations. In particular,
they will permit a look at 
aspects such as variability at the rotational and seasonal timescales [17, 18, 37] 
or the north-south brightness asymmetry [38].

\newpage
\textbf{Radiative transfer model}\\

The radiative transfer calculations were carried out with a backward Monte Carlo 
algorithm that has been described and thoroughly tested [20] and used to 
explore the information content in 
the diurnal and phase brightness modulations of Earth, Venus and exoplanets [29, 36, 39].
For the calculations presented here, we omitted the vector (polarization) treatment
of the equations by zeroing all terms of the scattering matrix except its (1, 1) entry.
\\

Each simulation was initiated by launching a photon from a randomly-selected location 
above Titan's disk, whether illuminated by direct sunlight or not (this is described in 
 \textit{Integration over the entire disk} [20]). 
Choosing a large enough integration disk 
ensures that the integration is not artificially truncated.
In the jargon of Monte Carlo photon simulations, each photon is assigned an initial
`life' of 1 at the beginning of the simulation. As the photon scatters in the optical
medium, its `life' progressively diminishes. The photon is considered to have
been fully absorbed by the medium when its left-over `life' falls below an established
threshold (10$^{-6}$ in our simulations). 
Because in the backward treatment of the scattering problem
the photon trajectory is independent of the 
location of the Sun, it is possible to conduct simulations with an arbitrary number of 
Suns simultaneously. In the simulations presented here, we considered 91 simultaneous Sun 
positions (from 0 to 180$^{\circ}$ in steps of 2$^{\circ}$), 
which allowed us to obtain the entire phase curve in a single simulation. 
Each phase curve was obtained by simulating one million photons.
\\

We further tested the radiative transfer model by comparing its output to published
calculations in stratified atmospheres sounded in limb viewing 
[40, 41]. 
The additional tests included $\sim$200 configurations: with/without aerosols; at three
different wavelengths; for tangential altitudes from 10 to 60 km; 
with/without considering polarization effects. 
The differences between the calculations from
another Monte Carlo model (MCC++ [41]) and our model were consistently
$<$1\%.

\newpage
\textbf{Atmospheric model}\\

The radiative transfer calculations done with the Monte Carlo algorithm take as inputs: 
\begin{itemize}
\item For the atmosphere: vertical profiles of 
extinction coefficients, $\gamma$($z$), 
single scattering albedos, $\omega_{0}$($z$), 
and scattering phase functions, $p$($\theta$)($z$).  
\item For the surface: surface reflectances, $r_{\rm{s}}$.
\end{itemize}
Each of these properties ($\gamma$, $\omega_{0}$, $p$($\theta$), $r_{\rm{s}}$) depends on 
the specific wavelength of the simulated photons.
The profiles of $\gamma$, $\omega_{0}$, and $p$($\theta$) are represented in the 
radiative transfer model over a total of 250 vertical slabs, each of them 2-km thick. 
\\ 

Our baseline description of the aerosol optical properties 
is based on the latest revision of the DISR observations [16]. 
The implemented properties were interpolated/extrapolated 
in wavelength from the DISR prescriptions [15, 16]
to the corresponding filter effective wavelengths. 
The DISR optical properties are well constrained between 490 and 950 nm 
for $\gamma_{\rm{a}}$ and $\omega_{0,\rm{a}}$ [16], 
and down to 355 nm for $p_{\rm{a}}$($\theta$)
[15, 16]. 
Model calculations based on the DISR implementation [16]
have proven consistent with past spectroscopic measurements of the geometric albedo 
between 500 and 950 nm [14], a fact that is confirmed in this work. 
Indeed, the solid red curves of Fig. 1, 
which are based exclusively 
on the DISR aerosol implementation plus Rayleigh scattering by the N$_2$/CH$_4$ gas 
(but without methane absorption), 
reproduce well the measured phase curves over the available range of phase angles. 
For these filter combinations 
(CL1$\_$BL2,
BL1$\_$CL2,
CL1$\_$GRN,
CL1$\_$CB1,
CL1$\_$CB2,
CL1$\_$CB3), 
the `best fits' presented in Fig. 1 as solid red curves
are simply the outputs of the radiative transfer model without adjusting any input. 
\\

For the atmospheric gas (98.4\%--1.6\% by volume of N$_2$--CH$_4$), 
we determine the scattering coefficient $\beta_{\rm{g}}$ and
scattering phase function $p_{\rm{g}}$($\theta$) from the classical
expressions for Rayleigh scattering. 
Number densities of the gas are based on the HASI profiles of temperature and pressure [42]. 
Because the impact of Rayleigh scattering is relatively small, 
the uncertainties associated 
with the HASI profiles have a negligible effect on the calculations.
A number of ro-vibrational absorption bands of methane occur at visible
and near infrared wavelengths. 
In our radiative transfer calculations,
we parameterize the effect of methane absorption over both broad- and narrowband
filter combinations by invoking filter-specific single scattering albedos 
$\omega_{0,\rm{g}}$ such that $\gamma_{\rm{g}}$=$\beta_{\rm{g}}$/$\omega_{0,\rm{g}}$
and $\alpha_{\rm{g}}$=(1/$\omega_{0,\rm{g}}$$-$1)$\beta_{\rm{g}}$
for the corresponding gas extinction and absorption coefficients, respectively.
This simplified approach intends to reproduce the empirical phase curves affected by
methane absorption with a minimum number of free parameters, and does it by assuming
an altitude-independent treatment of methane absorption over each specific spectral passband.
\\ 
 
For the optical properties of the aerosol-plus-gas mixture, we follow the usual 
summation rules: $\alpha$=$\alpha_{\rm{a}}$+$\alpha_{\rm{g}}$, 
$\beta$=$\beta_{\rm{a}}$+$\beta_{\rm{g}}$, 
$\gamma$=$\gamma_{\rm{a}}$+$\gamma_{\rm{g}}$, and
$\beta$$p$($\theta$)=$\beta_{\rm{a}}$$p_{\rm{a}}$($\theta$)+$\beta_{\rm{g}}$$p_{\rm{g}}$($\theta$).
\\

By definition, 
the solid red curves of Fig. 1 described above assume 
$\omega_{0,\rm{g}}$=1, which means that in those passbands 
the gas scatters but does not absorb. 
In other passbands (CL1$\_$MT1, RED$\_$CL2, 
CL1$\_$CL2, CL1$\_$MT2, IR2$\_$CL2, 
CL1$\_$MT3, CL1$\_$IR3), methane does absorb, a fact that is readily confirmed from
Fig. 2.
In these specific passbands, we ran a battery of radiative transfer calculations to explore
the impact of $\omega_{\rm{0,g}}$($\leq$1) on the model phase curves. 
The best fits shown in Fig. 1 as dashed red curves were obtained by 
minimization of the relative difference ($\sigma$, quoted in the figure) 
between measurements and calculations. 
Values of $\omega_{\rm{0,g}}$$\sim$1 mean that little methane absorption must be 
introduced, whereas values $\ll$1 mean that methane absorption is very effective at 
controlling the amount of radiation exiting the atmosphere. 
The best-fitting values of $\omega_{\rm{0,g}}$ were: 
0.575 (CL1$\_$MT1), 
0.900 (RED$\_$CL2), 
0.500 (CL1$\_$CL2), 
0.075 (CL1$\_$MT2), 
0.200 (IR2$\_$CL2), 
0.005 (CL1$\_$MT3), 
and
0.090 (CL1$\_$IR3). 
These numbers correlate with the strength of the methane bands in the corresponding
filter passbands. 
\\

Initial calculations showed that the DISR aerosol implementation did not 
provide a good match to the measured phase curves for the UV2$\_$CL2 
($\lambda_{\rm{eff}}$=306 nm) and CL1$\_$UV3 (343 nm) filter combinations. 
This was expected as the effective wavelengths of these filters fall outside the
wavelength validity range for the DISR aerosol implementation. 
In the calculations for these two filter combinations, 
we adopted altitude-independent $p_{\rm{a}}$($\theta$) functions, 
which were calculated as in previous work [15, 22]. 
Keeping with the ideas that led from 
the original DISR aerosol prescription [15] to the latest version [16], 
we modified by an adjustable amount
the aerosol single scattering albedo $\omega_{0,\rm{a}}$ 
above 150 km with respect to the values obtained with the DISR
analytical expressions [16]. 
An iterative process in which we ran many radiative transfer calculations each of them
with a different correction to $\omega_{0,\rm{a}}$($z$$>$150 km) led to the best fits
shown with dashed red curves in Fig. 1. 
The actual corrections were:
$\omega_{0,\rm{a}}^{\rm{old}}$($z$$>$150 km)$\rightarrow$$\omega_{0,\rm{a}}^{\rm{new}}$=1.275$\omega_{0,\rm{a}}^{\rm{old}}$($z$$>$150 km)
(UV2$\_$CL2) and
$\omega_{0,\rm{a}}^{\rm{old}}$($z$$>$150 km)$\rightarrow$$\omega_{0,\rm{a}}^{\rm{new}}$=1.175$\omega_{0,\rm{a}}^{\rm{old}}$($z$$>$150 km)
(CL1$\_$UV3). Thus, the implemented aerosols above 150 km for these two filter combinations 
 absorb less than in the original DISR prescription.
\\

For the radiative transfer calculations, we assumed Lambert-like reflection
at Titan's surface.
The wavelength-dependent surface 
reflectances were borrowed from Table 1 of a relevant work [43].
Extrapolation from/interpolation between the tabulated values yielded the following
surface reflectances adopted in our calculations: 
0.023 (UV2$\_$CL2),
0.035 (CL1$\_$UV3),
0.067 (CL1$\_$BL2),
0.072 (BL1$\_$CL2),
0.109 (CL1$\_$GRN),
0.123 (CL1$\_$MT1),
0.123 (CL1$\_$CB1),
0.131 (RED$\_$CL2),
0.132 (CL1$\_$CL2),
0.144 (CL1$\_$MT2),
0.147 (CL1$\_$CB2),
0.151 (IR2$\_$CL2, CL1$\_$MT3, CL1$\_$IR3, CL1$\_$CB3).\\

\cleardoublepage
\newpage
\textbf{The solar energy deposition rate}\\

The solar energy deposition rate is the difference between the
rates for incident solar energy and energy scattered by Titan: 
$$P_{\rm{dep}}=P_{\rm{inc}}-P_{\rm{sca}}.$$

For the incident contribution, we used:
$$
P_{\rm{inc}}=\int_{200 \rm{\;nm}}^{5000 \rm{\;nm}} \pi R^2_{\rm{opt}}(\lambda) F_{\odot}(\lambda) d\lambda.
$$
Here, $F_{\odot}$ is the solar irradiance (scaled to a solar 
constant $S_{\odot}$=14.9 Wm$^{-2}$ for Saturn's semi-major axis of 9.58 AU). 
$R_{\rm{opt}}$ is the optical radius, or 
 \textit{the radius of a sphere blocking the same amount 
of light on a screen behind Titan} [44].
This definition is equivalent to that of the $z$($\tau_{\rm{eq}}$=0.56) altitude 
that is often found in exoplanet studies [45].
The definition omits refraction effects, which depend on the
observer's location [19, 46].
Because Titan is not perfectly spherical, 
$P_{\rm{inc}}$ must ideally be evaluated with an $R_{\rm{opt}}$ that
accounts for changes in the limb optical thickness along the terminator.
Choosing the valid $R_{\rm{opt}}$ is challenging because terminator-averaged
determinations do not always exist, in which cases $R_{\rm{opt}}$ must be estimated indirectly. 
Furthermore, $R_{\rm{opt}}$ may change over time as both the detached and 
main haze layers evolve [26]. 
Titan's optical radius could be measured at a given time 
if the moon was observed transiting the Sun or another luminous extended object in 
the sky, as has been done at X-ray wavelengths [47]. 
It could also be determined from 
spatially-resolved measurements of Titan's shadow on Saturn, followed by
integration along the terminator [44].
\\

We evaluated the incident solar energy  rate and associated uncertainties
(details on the numerical implementation below), and obtained 
$P_{\rm{inc}}$=(3.87$\pm$0.07)$\times$10$^{14}$ W. 
The uncertainty in $R_{\rm{opt}}$ contributes about a third to the uncertainty in $P_{\rm{dep}}$. \\

In turn, for the scattered contribution:
$$
P_{\rm{sca}}=\int_{200 \rm{\;nm}}^{2500 \rm{\;nm}} \pi R^2 A_{\rm{g}}(\lambda) q(\lambda) F_{\odot}(\lambda) d\lambda,
$$
where $R$ is again Titan's solid radius used in the phase curve normalization.
We evaluated this integral and the associated uncertainties  
(numerical details below), to 
obtain $P_{\rm{sca}}$=(1.03$\pm$0.08)$\times$10$^{14}$ W.\\

Interestingly, the uncertainties associated with $P_{\rm{inc}}$ and $P_{\rm{sca}}$ are 
comparable. 
The specified integration limits for $P_{\rm{inc}}$ and $P_{\rm{sca}}$
ensure convergence of the integrals. 
Figure 4 and
Supplementary Table 1 convey important information on
the evaluation of $P_{\rm{inc}}$ and $P_{\rm{sca}}$. 

We can estimate the uncertainties in $P_{\rm{inc}}$ and $P_{\rm{sca}}$ by differentiation:
$$
\delta P_{\rm{inc}}^{R_{\rm{opt}}}=\int 2\pi R_{\rm{opt}}(\lambda) \delta R_{\rm{opt}}(\lambda) F_{\odot}(\lambda) d\lambda
$$
$$
\delta P_{\rm{sca}}= \left[ (\delta P_{\rm{sca}}^{A_{\rm{g}}})^{2}+(\delta P_{\rm{sca}}^{q})^{2} \right]^{1/2}
$$
with
$$
\delta P_{\rm{sca}}^{A_{\rm{g}}}
=\int \pi R^2_{\rm{p}} \delta A_{\rm{g}}(\lambda) q(\lambda) F_{\odot}(\lambda) d\lambda 
$$
$$
\delta P_{\rm{sca}}^{q}
=\int \pi R^2_{\rm{p}} A_{\rm{g}}(\lambda) \delta q(\lambda) F_{\odot}(\lambda) d\lambda.
$$

For the numerical evaluation of all the above integrals, we adopted:
\begin{itemize} 
\item Solar irradiance $F_{\odot}$($\lambda$) [48] 
scaled to the solar constant $S_{\odot}$=$\int F_{\odot}(\lambda)d\lambda$=14.9 Wm$^{-2}$.

\item Optical radius, $R_{\rm{opt}}$($\lambda$). For $\lambda$$<$1050 nm,
we estimated $R_{\rm{opt}}$($\lambda$) from the DISR aerosol implementation 
[15] and the expression for continuum absorption in exponential atmospheres
that establishes that $R_{\rm{opt}}$($\lambda$) is
equivalent to the altitude where the limb optical thickness $\tau_{\rm{eq}}=$0.56
[44, 45]. 
For $\lambda$$>$1050 nm, we adopted $R_{\rm{opt}}$($\lambda$)=2575 km + $z_{\rm{eff}}$, 
with $z_{\rm{eff}}$ as empirically inferred [49]. We matched the 
$R_{\rm{opt}}$($\lambda$) prescriptions for 
$\lambda$$<$1050 nm and $\lambda$$>$1050 nm by shifting the latter upwards
by 16 km. 

It is difficult to estimate the uncertainty 
$\delta R_{\rm{opt}}(\lambda)$, but given the 
small offset of 16 km between our $R_{\rm{opt}}(\lambda)$ predictions at 1050 nm from the
shorter- and longer-wavelength parameterizations, the achromatic 
choice $\delta R_{\rm{opt}}(\lambda)$=25 km is a valid guess. 
This is supported by measurements 
of the optical limb altitude along the terminator [44], which
varies by less than a scale height at wavelengths close to the peak 
of the solar spectrum ($\sim$500 nm), 
but by up to three scale heights in the near infrared ($\sim$940 nm).
The fact that the impact of $R_{\rm{opt}}(\lambda)$ on $P_{\rm{inc}}$ is largest near 500 nm
(Fig. 4a) provides additional support to our choice for $\delta R_{\rm{opt}}(\lambda)$. 
$R_{\rm{opt}}$($\lambda$) may evolve over time in response to changing irradiation
conditions, a possibility whose consequences are difficult to quantify.

\item Geometric albedo, $A_{\rm{g}}$($\lambda$). 
The good match between the ground-based spectroscopic measurements [14] and
our Cassini/ISS determinations 
(Fig. 2) gives confidence in the consistency of 
both datasets. 
Thus, in the evaluation of 
$P_{\rm{sca}}$ between 300 and 1050 nm, we adopted the spectrally-resolved
$A_{\rm{g}}$($\lambda$) data [14] multiplied by 1.02. Our models show that 
1.02 is a reasonable correction for the phase law $\Phi$($\alpha$) from $\alpha$=5.7$^{\circ}$ to
$\alpha$=0. This correction factor is consistent with the solar phase angle variation
reported in a multi-decade ground-based photometric investigation [37].
Shortwards of 300 nm, we adopted $A_{\rm{g}}$($\lambda$) values reported in an
ultraviolet investigation of Titan [50] 
and re-scaled them by 1.20 to match the longer-wavelength measurements at 300 nm [14].
For $\lambda$$>$1050 nm and $<$2500 nm, we adopted $A_{\rm{g}}$($\lambda$) values 
based on ground-based observations ([18]; their Fig. 5), after 
properly correcting them to our normalization radius. 

For $\lambda$$<$1050 nm the geometric albedos are likely accurate to a few percent,
and we assume $\delta A_{\rm{g}}$/$A_{\rm{g}}$$\sim$5\%. 
For $\lambda$$>$1050 nm, rotational, seasonal and secular changes in Titan's 
reflectance may easily cause $\delta A_{\rm{g}}$/$A_{\rm{g}}$$\sim$10\% [18].

\item Phase integral, $q$($\lambda$). 
The broadband filter combinations 
UV2$\_$CL2, 
CL1$\_$UV3, 
BL1$\_$CL2, 
CL1$\_$GRN, 
RED$\_$CL2, 
IR2$\_$CL2, and CL1$\_$IR3
provide nearly-continuous, passband-integrated insight into the phase integral up to $\sim$1000 nm. 
This information was incorporated into the evaluation of $P_{\rm{sca}}$ by
building a `continuous' $q_{\rm{fit}}$($\lambda$) that fits the corresponding pairs of 
$\lambda_{\rm{eff}}$ and $q$($\lambda_{\rm{eff}}$) between 306 and 928 nm
(Fig. 4). 
The uncertainties in the phase integral over this spectral range are due to: 
the loss of detail introduced by the replacement of the true 
$q$($\lambda$) by $q_{\rm{fit}}$($\lambda$); the unconfirmed shape of the Titan phase
curves for phase angles $\alpha$$>$166$^{\circ}$ unobserved by Cassini/ISS. 
From the comparison of the $q$ values specific to broadband and narrowband filters, and
the fact that phase angles $\alpha$$>$166$^{\circ}$ typically 
contribute $\lesssim$5 \% to the phase integral, we suggest 
an average uncertainty $\delta q$/$q$$\sim$5\% for $\lambda$$<$1050 nm.

Outside the Cassini/ISS spectral range, 
we adopted $q$($\lambda$$<$306 nm)=$q$($\lambda$=306 nm) and $q$($\lambda$$>$1050 nm)=2.25. 
The specific value of $q$($\lambda$) at the shorter wavelengths is not critical because 
both the solar output and Titan's overall reflectance are small at these wavelengths, and 
the contribution of these wavelengths to $P_{\rm{sca}}$ is also small.
For $\lambda$$>$1050 nm, based on our findings at shorter wavelengths, 
we assumed that $q$ is in the range between 2 and 2.5, and therefore
$\delta q$/$q$$\sim$10\%

\end{itemize}

\newpage

\cleardoublepage
\textbf{Data availability statement}\\

The data that support the plots within this paper and other findings of this study are
available from the corresponding author upon reasonable request.\\

\textbf{Additional references}\\

[35] Knowles, B.
Cassini Imaging Science Subsystem (ISS). Data User's Guide. 
Cassini Imaging Central Laboratory for Operations (CICLOPS)
(2014).\\

[36] Garc\'ia Mu\~noz, A.
Towards a comprehensive model of Earth's disk-integrated Stokes vector. 
\textit{Int. J. Astrobiology}, 
\textbf{14}, 379--390 
(2015).
\\

[37] Lockwood, G.W. \&
Thompson, D.T.
Seasonal photometric variability of Titan, 1972-2006.
\textit{Icarus}, \textbf{200}, 616--626
(2009).
\\

[38] Sromovsky, L.A., 
Suomi, V.E., 
Pollack, J.B., 
Krauss, R.J., 
Limaye, S.S., et al. 
Implications of Titan's north-south brightness asymmetry. 
\textit{Nature}, \textbf{292}, 698--702
(1981).
\\

[39] Garc\'ia Mu\~noz, A., 
P\'erez-Hoyos, S. \& 
S\'anchez-Lavega, A.
Glory revealed in disk-integrated photometry of Venus. 
\textit{Astron. \& Astrophys.}, \textbf{566}, id.L1
(2014).
\\

[40] Loughman, R.P., 
Griffioen, E., 
Oikarinen, L., 
Postylyakov, O.V., 
Rozanov, A., 
Flittner, D.E., 
Rault, D. F.
Comparison of radiative transfer models for limb-viewing scattered sunlight measurements.
\textit{J. Geophys. Res. -- Atmos.}, 
\textbf{109}, D06303
(2004).\\

[41] Postylyakov, O.V.
Linearized vector radiative transfer model MCC++ for a spherical atmosphere.
\textit{J. Quant. Spectrosc. R.A.}, \textbf{88}, 297--317
(2004).
\\

[42] Fulchignoni, M., 
Ferri, F., 
Angrilli, F., 
Ball, A.J., 
Bar-Nun, A., et al.
In situ measurements of the physical characteristics of Titan's environment.
\textit{Nature}, \textbf{438}, 785--791 
(2005).
\\

[43] Karkoschka, E. \& Schr\"oder, S.E.
Eight-color maps of Titan's surface from spectroscopy with Huygens' DISR
\textit{Icarus}, \textbf{270}, 260--271
(2016).\\

[44] Karkoschka, E. \& Lorenz, R.D.
Latitudinal variation of aerosol sizes inferred from Titan's shadow.
\textit{Icarus}, \textbf{125}, 369--379
(1997).\\

[45] Lecavelier des Etangs, A.,
Vidal-Madjar, A.,
D\'esert, J.-M. \&
Sing, D.
Rayleigh scattering by H$_2$ in the extrasolar planet HD 209458b.
\textit{Astron. \& Astrophys.}, \textbf{485}, 865--869
(2008).
\\

[46] Garc\'ia Mu\~noz, A., 
Zapatero Osorio, M.R., 
Barrena, R., 
Monta\~n\'es-Rodr\'iguez, P., 
Mart\'in, E.L. \& Pall\'e, E.
Glancing views of the Earth: From a lunar eclipse to an exoplanetary Transit.
\textit{Astrophys. J.}, \textbf{755}, id.103 
(2012).
\\

[47] Mori, K.,
Tsunemi, H.,
Katayama, H., 
Burrows, D.N., 
Garmire, G.P. \& Metzger, A.E.
An X-Ray measurement of Titan's atmospheric extent from its transit of the Crab Nebula.
\textit{Astrophys. J.}, \textbf{607}, 1065--1069
(2004).
\\

[48] Wehrli, C. 
Extraterrestrial Solar Spectrum. \textit{PMOD publication 615}
(1985).
\\

[49] Robinson, T.D., 
Maltagliati, L., 
Marley, M.S. \& 
Fortney, J.J.
Titan solar occultation observations reveal transit spectra of a hazy world.
\textit{PNAS}, \textbf{111}, 9042--9047
(2014).
\\

[50] McGrath, M.A.,
Courtin, R.,
Smith, T.E.,
Feldman, P.D. \&
Strobel, D.F.
The ultraviolet albedo of Titan.
\textit{Icarus}, \textbf{131}, 382--392
(1998).
\\

\end{document}